\documentclass{aa}
\usepackage{graphicx}
\def\approxgt{\mathrel{\hbox{\rlap{\lower.55ex \hbox {$\sim$}}
        \kern-.3em \raise.4ex \hbox{$>$}}}}
\def\approxlt{\mathrel{\hbox{\rlap{\lower.55ex \hbox {$\sim$}}
        \kern-.3em \raise.4ex \hbox{$<$}}}}
\begin{document}
   \title{The history of the iron K$_{\alpha}$ line profile \\ in the
Piccinotti AGN ESO198-G24}

   \author{Matteo Guainazzi
          \inst{1}
          }

   \offprints{M.Guainazzi}

   \institute{XMM-Newton Science Operation Center, VILSPA, ESA, Apartado
              50727,E-28080 Madrid, Spain \\
              \email{mguainaz@xmm.vilspa.esa.es}
              }

   \date{Received 26 August 2002; accepted 30 January 2003}

   \abstract{
This paper presents ASCA (July 1997),
XMM-Newton (December 2000) and BeppoSAX
(January 2001) observations of the Piccinotti Seyfert~1 galaxy
ESO198-G24.
The BeppoSAX 0.1--200~keV spectrum exhibits reprocessing features,
probably produced by an X-ray illuminated,
relativistic accretion disk subtending a solid angle
$\approxlt 2 \pi$. During the XMM-Newton observation
the fluorescent iron K$_{\alpha}$ line
profile (centroid energy
$E_{\rm c} \simeq 6.4$~keV)
was broad and twice as bright as in the BeppoSAX observation.
An additional emission feature ($E_{\rm c} \simeq 5.7$~keV),
detected at the 96.3\% confidence level, may be part
of a relativistic, double-peaked profile.
By contrast, in the earlier ASCA observation
the line profile is
dominated by a remarkably narrow ``core''
(intrinsic width, $\sigma < 50$~eV). If
this component is produced by
reflection off the inner surface of a molecular torus, its
large Equivalent Width ($\simeq 300$~eV) most likely represents
the ``echo'' of a previously brighter flux state,
in agreement with the dynamical
range covered by the historical 
X-ray light curve in ESO198-G24.

	}
\authorrunning{Guainazzi}

\titlerunning{Iron K$_{\alpha}$ line in ESO198-G24}

   \maketitle
%

\section{Introduction}

The discovery of the broadened and skewed
fluorescent iron K$_{\alpha}$ line profile
in the first ``long-look''
ASCA observation of MCG-6-30-15 (\cite{tanaka95})
fostered the hope that general relativistic effects
could be observationally studied through X-ray
spectroscopy of nearby Active Galactic Nuclei (AGN).
This hope was 
further strengthened by the discovery that about 50\% of
the known bright Seyfert~1s exhibit broad line
profiles (\cite{nandra97}). It was reasonable to expect that
the advent of {\it Chandra} and XMM-Newton would
allow to significantly extend the depth and range
of these studies.

These hopes are now facing a complex reality.
In a few objects the existence of a relativistically broadened
iron line profile is out of question (\cite{nandra99},
\cite{wilms01},
\cite{turner02}, \cite{fabian02}). However,
often the observed profiles do not match
the theoretical calculations (\cite{fabian89};
\cite{laor91}; \cite{matt92}). Electron
scattering, as often observed in type~2
Seyferts (\cite{ueno94}; \cite{turner97};
\cite{matt00}), may help reconciling the difference
between observed and theoretical
line profiles produced in X-ray illuminated,
relativistic accretion disks
(\cite{reeves01}; \cite{matt01}).
Other pieces of evidence
suggest constraints on the nature of
the accretion: the disk may be truncated at a
radius where relativistically broadened
wings are negligible (see, e.g. the
discussion in \cite{obrien01}), or
may develop a hot ``skin''
(\cite{nayakshin00}, \cite{ballantyne01}),
which may contribute to the iron line profile
through transitions of highly ionized
stages.
Observations with the {\it Chandra}
high-energy gratings (\cite{yaqoob01}, \cite{kaspi02},
\cite{weaver01}, and references therein)
and with XMM-Newton (\cite{gondoin01a}; \cite{gondoin01b},
\cite{pounds01}; \cite{petrucci02})
have often unveiled narrow line components.
When measured at the currently highest possible
energy resolution, 
their intrinsic velocities are consistent with
the width of the optical broad lines,
suggesting an origin in the same medium.
However, different origins
(e.g.: scattering off the inner side of
the molecular ``torus''
envisaged by the Seyfert unification scenarios; \cite{antonucci85};
\cite{antonucci93}) cannot be ruled out.
These narrow components have Equivalent Widths
(EW) typically
as large as $100$~eV, and may contribute as much as
50\% to the total iron line flux
(\cite{weaver01}).

In this context, we present in this
paper X-ray observations
of the Seyfert~1 galaxy ESO198-G24 ($z = 0.0455$).
Relatively little is known on its X-ray
spectral properties, despite the fact that it
belongs to the
Piccinotti sample (H0235-52;
\cite{piccinotti82}), and hence it is one of
the brightest AGN of the 2--10~keV
sky. In their EXOSAT AGN survey paper, Turner \& Pounds
(1989) note that it exhibits a ``canonical spectrum''
(photon index, $\Gamma = 1.88 \pm^{0.16}_{0.08}$)
with ``no significant low-energy absorption'', and
a 2--10~keV luminosity $\simeq 1.64 \times
10^{44}$~erg~s$^{-1}$, about three times lower than
during the HEAO-1 scans. In the ROSAT
All-Sky Survey (\cite{schartel97}) a rather steeper
soft spectral index was found ($\Gamma = 2.5 \pm 0.1$),
again without any evidence for absorption. The
0.1--2.4~keV observed flux 
[$(5.33 \pm 0.13) \times 10^{-11}$~erg~cm$^{-2}$~s$^{-1}$]
corresponds to a luminosity of
$\simeq 2.0 \times 10^{44}$~erg~s$^{-1}$
once corrected for Galactic absorption
($N_{\rm H,Gal} = 3.2 \times 10^{20}$~cm$^{-2}$;
\cite{dickey90}) and extrapolated into the 2--10~keV band.
Malizia et al. (1999) report a detection by
BATSE, with a 2--100~keV flux of
$(5.3 \pm 1.7) \times 10^{-11}$~erg~cm$^{-2}$~s$^{-1}$.
To our knowledge, no result has ever been published from
ASCA, BeppoSAX or XMM-Newton observations of this source so
far.

This paper attempts to address
this lack in the literature.
The log of the observations presented in this paper
is reported
in Table~\ref{tab4}, together with the
corresponding exposure times and count rates
\begin{table*}
\caption{Log of the observations presented in this paper}
\begin{center}
\begin{tabular}{lccc} \hline \hline
Mission & Observation date & Exposure time & Count rates \\
& & (ks) & (s$^{-1}$) \\ \hline 
ASCA (SIS0/GIS2) & July 10, 1997 & 31.1/34.3 & $0.153 \pm
0.003$/$0.135\pm 0.002$ \\
XMM-Newton (pn/RGS1) & December 1, 2000 & 6.8/13.0 & $5.50 \pm
0.03$/$0.184 \pm 0.003$ \\
BeppoSAX (LECS/MECS/PDS) & January 23, 2001 & 55.1/143.3/51.3 & $0.1048 \pm
0.0015$/$0.1723 \pm 0.0011$/$0.32 \pm 0.03$ \\ \hline \hline
\end{tabular}
\end{center}
\label{tab4}
\end{table*}
in the energy bands where the spectral analysis discussed
in Sect.~2 was carried out.
The paper
is organized as follows:
Sect.~2 describes the spectral analysis, employing
phenomenological models only.
A comparison of the observed
fluorescent iron K$_{\alpha}$ line profiles with the predictions
of theoretical models is presented in Sect.~3.
The results are discussed in Sect.~4.
Throughout this paper: energies are quoted in
the source reference frame;
uncertainties are at the 90\% confidence level for one
interesting parameter; $H_{\rm 0} = 50$~km~s$^{-1}$~Mpc$^{-1}$
and $q_{\rm 0} = 0.5$, unless otherwise specified.

\section{Spectral results}

\subsection{XMM-Newton}

In this paper only the time-averaged
pn (\cite{struder01}) spectrum in the 2--15~keV band will be
presented.
No significant spectral variability
is associated with a $\pm 7\%$
flux fluctuation observed during the
XMM-Newton observation.
The source was outside the MOS1 field of view.
The MOS2 exposure was performed
in Timing Mode. The calibration of
this mode is still preliminary, and the corresponding data
will not de discussed in this paper.
The pn
observation was performed in Small Window Mode with the
blocking optical
Medium filter. Data were reduced with {\sc Sas v5.3.3}
(\cite{jansen01}),
using the most updated calibration files
available as of July 1, 2002. A spectrum including
single- and double-pixel events was extracted. It was
verified that
spectra extracted with either only single- or only double-pixel
events are mutually consistent within the statistical uncertainties.
Source spectra were extracted from a circular region of
1.8$\arcmin$ radius.
Non X-ray background remained low throughout the observation,
hence the
whole integration time of the observation was used.
Several recipes to extract
the background spectra were compared,
and yielded very similar
results.
The results presented in this paper
were obtained with
background spectra
extracted from
blank field templates available at the XMM-Newton
Science Operation Center (\cite{lumb02}), and rescaled
in order to match the observed 
spectrum in the 15--20~keV energy band.
The background contributes around 3\% and
8\% of the total spectrum at 2 and 5.5~keV respectively,
and is brighter than the source above 10~keV. The spectra
used 
throughout this paper were rebinned
in such a way that: a) the intrinsic
energy resolution of the detectors is sampled
by a number of spectral channels not larger than
3; b) each spectral bin has at least 50
counts,
to ensure the applicability
of the $\chi^2$ test. The same criteria
were applied for the BeppoSAX and ASCA spectra discussed
later, but the number of counts in these cases was
limited to 25.

All the models discussed in this Section and
in the following ones are
modified by photoelectric absorption, whose
column density, $N_{\rm H}$, is
constrained to be not lower than the contribution
due to the interstellar matter in our Galaxy along the line-of-sight
to ESO198-G24.

\begin{figure}
   \centering
   \includegraphics[angle=-90,width=8cm]{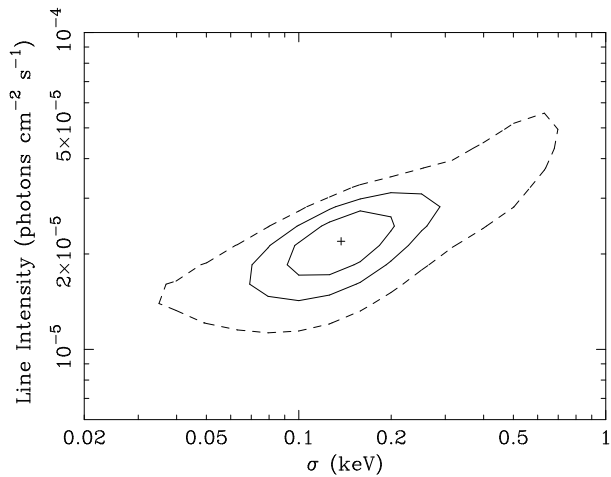}
      \caption{Iso-$\chi^2$ contours for the 
		fluorescent iron K$_{\alpha}$
		line intrinsic width ($\sigma$) versus
		intensity in the XMM-Newton  observation
		of ESO198-G24. Contours correspond to
		$\Delta \chi^2 = 1$, 2.71 ({\it
		solid lines}) and 4.61 ({\it
		dashed line}), respectively.
              }
         \label{fig2}
   \end{figure}

A simple power-law model is a marginally acceptable
description of the pn spectrum 
($\chi^2 = 146.5/114$ degrees of freedom, dof).
A count excess around 6~keV (observer's frame) is present. The
addition of a narrow (i.e. intrinsic width, $\sigma$, equal to 0)
Gaussian emission profile to the best-fit power-law yields
an improvement of the $\chi^2$ by 18.3
for a decrease by 2 in the number of degrees of
freedom (this quantity will be indicated
as $\Delta \chi^2/\Delta \nu$
hereinafter), significant at the 
99.992\% confidence level. Leaving the intrinsic width
of the Gaussian profile free in the
fit yields a further improvement in its
quality ($\Delta \chi^2/\Delta \nu = 5.8/1$,
significant at the 98.8\% confidence level).
The best-fit parameters of the line are:
$\sigma = 140 \pm^{180}_{80}$~eV,
centroid energy $E_{\rm c} = 6.43 \pm 0.07$~keV, and $EW = 180 \pm 70$~eV.
The centroid
energy is consistent with
K$_{\alpha}$ fluorescence from neutral or mildly
ionized iron, and strictly inconsistent with
species more ionized than Fe{\sc xix}. 
The intrinsic width is constrained to be larger than
about 40~eV at the 90\% confidence level for two
interesting parameters (see Fig.\ref{fig2}). The upper
limit is more loosely constrained, and widths as large as
0.5~keV are in principle possible.

\begin{figure}
   \centering
   \includegraphics[angle=-90,width=8cm]{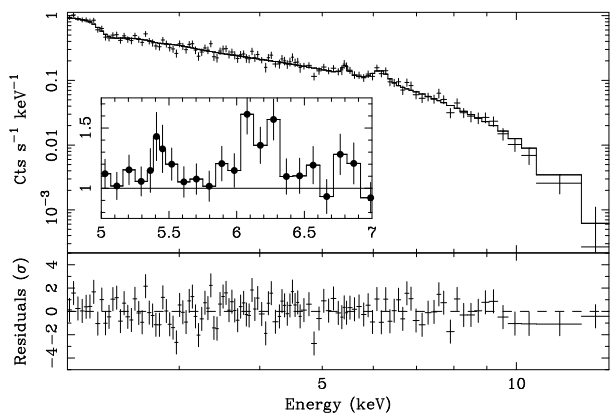}
      \caption{Spectrum ({\it upper panel}) and
		residuals in units of standard deviations
		({\it lower panel}) when the best-fit
		model as in Table~\ref{tab1}
		is applied to the pn data of ESO198-G24.
	        In the {\it inset}, the data/model
		ratio in the 5-7~keV energy range is shown,
		when the best-fit continuum is determined
		after the data points in the 5.0-6.5~keV
		(observer's frame) are removed.
              }
         \label{fig3}
   \end{figure}

\begin{figure}
   \centering
   \includegraphics[width=8cm]{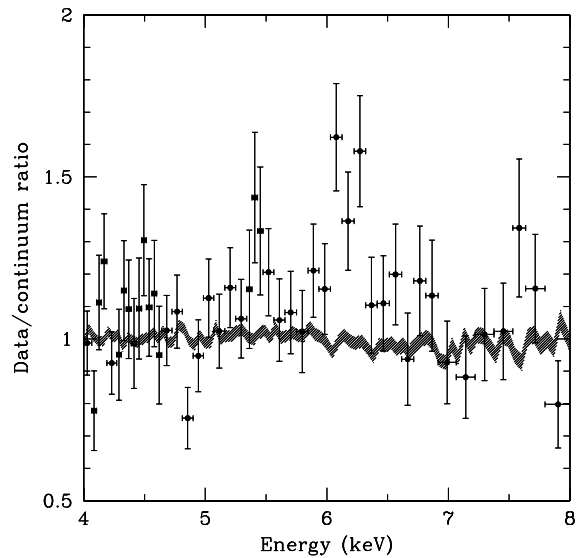}
      \caption{Ratio of the ESO198-G24 pn spectrum
		against the best-fit continuum model
		as in Table~\ref{tab1}
		({\it filled circles}). The {\it	
		shaded area} represents the $\pm 1$$\sigma$
		residuals
		against the best-fit 2--15~keV power-law model
		in an
		observation of the
		featureless AGN 3C273 (Obs.ID
		\#0136550101 in revolution 277).
		They represent	
		typical systematics associated with
		the pn response matrix employed in this
		paper.
              }
         \label{fig5}
   \end{figure}

The profile of the iron line may be more complex
than a single broad Gaussian.
A careful inspection of the residuals (see Fig.~\ref{fig3})
suggests an additional emission
feature with $E_{\rm c} \simeq 5.4$~keV
(observer's frame).
If an additional Gaussian profile
is added to the model, the
quality of the fit
is improved at the  96.3\% confidence level
($\Delta \chi^2/\Delta \nu = 
9.3/3$).
The best-fit values of this component are
$E_{\rm c} = 5.70 \pm^{0.07}_{0.12}$~keV,
$EW = 70 \pm 40$~eV and $\sigma < 280$~eV.
In order to compare its significance with the
current systematic uncertainties of the pn response matrix, we
analyzed pn public data of featureless calibration sources,
acquired in Small Window Mode and
reduced under the same conditions as the
ESO198-G24 data. Typical systematic
uncertainties (represented by the shaded area in
Fig.~\ref{fig5}) are $\pm 5\%$ in the 5--7~keV
band. We rule out therefore an instrumental origin
for the $E_{\rm c} \simeq 5.7$~keV feature.
Table~\ref{tab1} summarizes the spectral
\begin{table}
\caption{Best-fit parameters and results, when the iron line
emission is modeled with a combination of Gaussian profiles.
Continua are power-laws for the XMM-Newton and ASCA
observations, and a power-law modified by Compton-reflection
for the BeppoSAX observation. The subscripts ``SG'' and
``DG'' refers to the fit of the iron line profile
with a single or a double Gaussian profile, respectively}
\begin{center}
\begin{tabular}{lccc} \hline \hline
&  BeppoSAX & XMM-Newton & ASCA \\ \hline
\multicolumn{4}{l}{Continuum} \\
$F_{\rm h}$$^a$ & $15.04 \pm 0.10$ & $10.90 \pm 0.11$  & $5.91 \pm 0.16$ \\
$L_{\rm X}$$^b$ & $6.85 \pm 0.09$ & $4.20 \pm 0.09$  & $2.01 \pm 0.07$ \\ 
$\Gamma$ & $1.79 \pm 0.04$  & $1.77 \pm 0.03$ & $1.75
\pm^{0.05}_{0.03}$ \\
$N_{\rm H,int}$$^c$ & $<0.5$ & $\le 3.2$ & $\le 13.2$ \\ \hline
\multicolumn{4}{l}{``6.4~keV'' feature} \\
$E_{\rm c}$~(keV) & $6.4 \pm 0.2$  & $6.43 \pm 0.07$ & $6.40
\pm^{0.09}_{0.05}$ \\
$\sigma$~(eV) & $300 \pm 300$ &  $140 \pm^{120}_{70}$ & $0 \pm^{50}_0$
\\
$I$$^d$ & $1.7 \pm^{0.5}_{1.0}$  & $2.3 \pm^{1.2}_{0.7}$ & $2.2
\pm^{0.5}_{0.7}$ \\ 
$EW$~(keV) & $100 \pm^{30}_{60}$ & $190 \pm^{100}_{60}$ & $320
\pm^{70}_{100}$ \\ \hline
\multicolumn{4}{l}{``5.7~keV'' feature} \\
$E_{\rm c}$~(keV) & ... & $5.70 \pm^{0.07}_{0.12}$ & ... \\
$\sigma$~(eV) & ... & $0 \pm^{280}_0$ & ... \\
$I$$^d$ & ...  & $1.0 \pm 0.5$ & ... \\ 
$EW$~(keV) & ... & $70 \pm 40$ & ... \\ \hline 
$\chi^2_{\nu}|_{SG}$ & 1.14 & 1.11 & 1.06 \\ 
$\chi^2_{\nu}|_{DG}$ & ... & 1.06 & ... \\ \hline \hline
\end{tabular}
\end{center}

\noindent
$^a$2--10~keV observed flux in units of
$10^{-12}$~erg~cm$^{-2}$~s$^{-1}$

\noindent
$^b$Unabsorbed 0.1--100~keV luminosity in
units of $10^{44}$~erg~s$^{-1}$

\noindent
$^c$column density of the
intrinsic (i.e. at $z = z_{ESO198-G24}$) absorber
in units of $10^{20}$~cm$^{-2}$

\noindent
$^d$in units of $10^{-5}$~photons~cm$^{-2}$~s$^{-1}$

\label{tab1}
\end{table} 
results.

The
width of the $E_c \simeq 6.4$~keV feature is not
due to the presence of the additional line
at $E_c \simeq 5.7$~keV. If the data are
fit with
a model constituted by a power-law
and one Gaussian profile,
after removing the data points in the
5.0-5.5~keV energy range (observer's
frame), the parameters of the Gaussian
profile are: $E_c = 6.43 \pm 0.08$~keV,
$\sigma = 150 \pm^{180}_{70}$~eV,
$EW = 200 \pm^{100}_{80}$~eV,
therefore indistinguishable from those
obtained with the complete model on
the whole 2--15~keV energy band.

No further lines are
statistically required by the fit.
No evidence exists for photoelectric
absorption edges in the pn spectrum either. The 90\% upper 
limits on the optical depth of un-blurred
Fe{\sc i},
Fe{\sc xxv} and Fe{\sc xxvi} K edges are 0.14, 0.24 and 0.30,
respectively.

\subsection{ASCA}

ASCA data were retrieved from the public HEASARC archive as
screened event lists. Data reduction followed standard
procedures as described in, e.g., Guainazzi et al. (2000).
Spectra of all the ASCA instruments were integrated on the
whole elapsed time
of the observation, after verifying that one can neglect
spectral variability effects, and were
simultaneously
fit, allowing free normalization factors for
the individual instruments to
account for systematic uncertainties in the
cross-normalizations ($\pm 7\%$). A simple power-law
continuum modified by intervening photoelectric absorption
yields a marginally acceptable
fit to the data ($\chi^2 = 413.7/364$~dof).
The 2--10~keV flux is about two times weaker than in the
XMM-Newton observation.
The residuals exhibit a clear narrow excess
feature at about 6~keV (observer's frame). The addition of a narrow
Gaussian profile largely improves the quality
of the fit ($\Delta \chi^2/\Delta \nu = 28.7/2$,
significant at the 99.9998\% confidence level). 
The centroid energy is again consistent with neutral or
mildly ionized iron ($E_c = 6.40 \pm^{0.09}_{0.05}$).
No additional improvement in the $\chi^2$ is obtained
if the line width is left free in the fit. The 90\% upper limit on its
$\sigma$ is 50~eV only. Similarly, the inclusion of a
further Gaussian emission profile with $E_c \simeq 5.7$~keV
is not statistically required
($\Delta \chi^2/\Delta \nu = 4.2/3$).

\begin{figure}
   \centering
   \includegraphics[angle=-90,width=8cm]{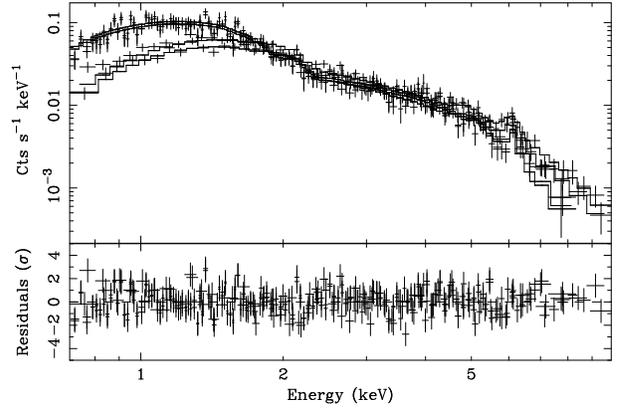}
      \caption{Spectrum ({\it upper panel}) and
		residuals in units of standard deviations
		({\it lower panel}) when the best-fit
		model as in Table~\ref{tab1}
		is applied to the ASCA data of ESO198-G24
              }
         \label{fig8}
   \end{figure}

A summary of the best-fit results is shown in Table~\ref{tab1}.
The ASCA spectra and corresponding best-fit model are shown
in Fig.~\ref{fig8}.

\subsection{BeppoSAX}

BeppoSAX data were retrieved from the
A.S.I. BeppoSAX Science Data Center (ASDC)
public archive. Spectra
from the Low Energy Concentrator Spectrometer (LECS,
0.1--4~keV; \cite{parmar97}) and the Medium Energy
Concentrator Spectrometer (MECS, 1.8--10~keV;
\cite{boella97b}) were extracted from linearized
and calibrated event lists, following standard recipes
as detailed in, e.g., Guainazzi et al. (1999).
Background spectra were extracted from blank
sky field event lists provided by the
ASDC, using the same area in detector coordinates
as the source.
Extraction
radius were 8$\arcmin$ and 4$\arcmin$
for the LECS and the MECS, respectively.
Background-subtracted spectra for the Phoswitch Detector System
(PDS, 13--200~keV; \cite{frontera97}) were generated by
plain subtraction of the 96~s duty-cycle intervals, when
the collimators were pointing to the line of sight towards
ESO198-G24 and to a region 3.5$^{\circ}$ degrees aside.
Spectra were integrated on the whole elapsed time
of the observation, after verifying that one can neglect
spectral variability effects,
and were simultaneously fit after applying correction
factors, to account for known differences in the absolute flux
cross-calibration of the instruments
(in particular, a value of 0.84
was employed for the PDS versus MECS normalization;
\cite{fiore98}).

\begin{figure*}
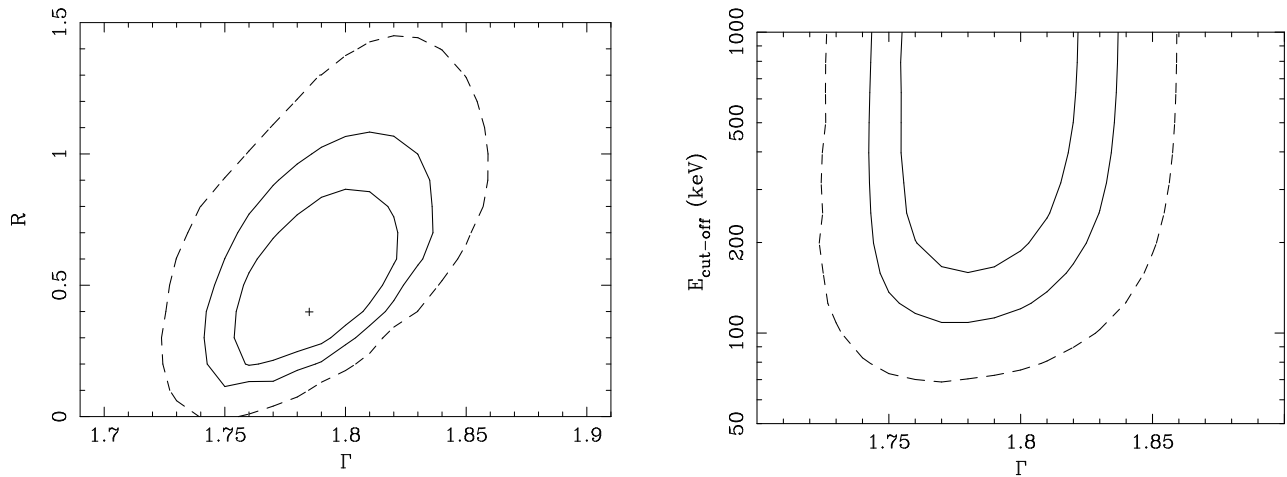

   \centering
   \hbox{
   \includegraphics[angle=-90,width=8cm]{H3946_F5a.ps}
   \hspace{0.75cm}
   \includegraphics[angle=-90,width=8cm]{H3946_F5b.ps}
   }
      \caption{Iso-$\chi^2$ contour plots for the
	      spectral index versus the Compton-reflection
	      $R$ parameter ({\it left panel}) and
	      the high energy cut-off ({\it
	      right panel}) for the BeppoSAX
	      observation of ESO198--G24. Contours
	      correspond to the 68\%, 90\% ({\it
	      solid lines}) and 99\% ({\it dashed lines})
              confidence
	      levels for two interesting parameters,
              respectively
              }
         \label{fig6}
   \end{figure*}

A marginally acceptable fit
($\chi^2 = 170.3/144$~dof) is obtained
with the standard Seyfert~1 continuum model in the BeppoSAX
energy bandpass (\cite{perola02}),
i.e., a photoelectrically absorbed
power-law, modified by Compton-reflection from cold
matter (\cite{lightman88}, \cite{george91}, \cite{magdziarz95}).
Assuming
an inclination angle of the reprocessing matter
$i = 25^{\circ}$, $R$\footnote{$R$ is proportional to the ratio
between the normalizations of the reflected and of
the primary continuum component. If the
primary emission is isotropic, $R = \Omega/2 \pi$,
where $\Omega$ is the solid angle subtended by
the reflector to the primary continuum source.}
is comprised
between 0.1 and 1.0 at the 90\% level for
two interesting parameters (cf. the left panel of
Fig.~\ref{fig6}).
Any thermal cut-off in the intrinsic power-law is constrained
to lay at energies $E_{\rm cutoff} \approxgt 150$~eV at
the same confidence level
(cf. the right panel of Fig.\ref{fig6}).
The spectral index of the {\it intrinsic} power-law
($\Gamma = 1.79 \pm 0.04$) is remarkably close
to the XMM-Newton and ASCA {\it observed}
indices.
The BeppoSAX best-fit parameters
values are only marginally affected
by different choices of the inclination angle,
e.g.:
$R$ changes by 0.1 and $\Gamma$ by 0.02
if $i = 45^{\circ}$.

\begin{figure}
   \centering
   \includegraphics[angle=-90,width=8cm]{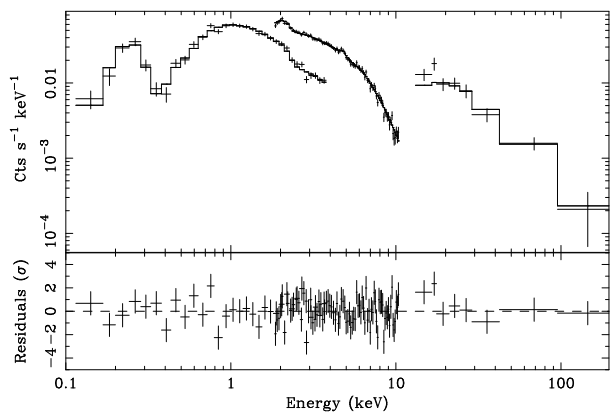}
      \caption{Spectrum ({\it upper panel}) and
		residuals in units of standard deviations
		({\it lower panel}) when the best-fit
		power-law plus Compton-reflection
		model
		is applied to the BeppoSAX data of ESO198-G24.
		The line is accounted for by a single broad
		Gaussian profile.
              }
         \label{fig7}
   \end{figure}

Excess residuals around $E \simeq 6$~keV (observer's frame)
again suggest emission from a fluorescent
iron K$_{\alpha}$ line. The addition of a Gaussian profile
yields an improvement in the quality of the fit
by $\Delta \chi^2/\Delta \nu = 8.5/3$, significant at
the 93.5\% confidence level.
No conclusion can be drawn on the intrinsic width
of the line profile.
Its total
EW does not exceed 130~eV. A summary of the
best-fit results is shown in Table~\ref{tab1}.
 The best-fit
model and residuals are shown in Fig.\ref{fig7}.

\section{The history of the iron line profile in ESO198-G24}

Table~\ref{tab1} summarizes the best-fit
results for the ASCA,
XMM-Newton and BeppoSAX observations of ESO198-G24,
when the iron line is
fit by a (combination of) Gaussian profile(s). These observations
span a range of about 3 in 2--10~keV X-ray flux, whereas
the spectral shape of the continuum does not exhibit
significant changes. Anecdotically, the ASCA observation
caught ESO198-G24 in its
faintest state ever reported. Some of the properties of the iron line
vary in a way, which is apparently correlated with the
X-ray flux. The line EW {\it decreases} with increasing
continuum flux, from $EW \simeq 320$~eV in the
weakest (ASCA)
to the $EW \simeq 100$~eV in the brightest (BeppoSAX)
state. The narrow profile
of the 6.4~keV emission line
measured by ASCA ($\sigma < 50$~eV)
is formally inconsistent with the broad profile
measured by XMM-Newton ($\sigma \simeq 140$~eV).
The centroid energy is always consistent with that of
fluorescent transitions of neutral or mildly ionized
iron. However, alternative interpretations are possible,
when the line profile is significantly broadened
or skewed.

Broad and asymmetric iron line profiles in AGN are naturally
explained as due to relativistic effects, affecting
the photons emitted in an X-ray illuminated accretion
disk (\cite{fabian89}, \cite{laor91}, \cite{matt92}). If the
line photons are emitted - originally with a monochromatic
energy distribution - within a few gravitational radii from
the source of an intense gravitational potential, the combination
of kinematic (Doppler) and gravitational shifts over a range
of disk annuli can produce significant broadening
and skewing of the profile.

The iron line profile can be best studied in the pn spectrum,
thanks to its combination of photon statistics and instrumental
energy resolution.
Bearing the above scenario in mind, the observed
excess in the 5--6.5~keV (observer's frame)
energy range was fit with
relativistic emission line profiles, as expected around a Schwarzschild
(model {\sc diskline} in {\sc Xspec}; \cite{fabian89}) or a Kerr
(model {\sc laor} in {\sc Xspec}; \cite{laor91}) black hole. These
models depend on a number of parameters: apart
from the centroid energy and the normalization,
the inner ($r_i$) and outer ($r_o$) radius of the
line emitting region in units of Schwarzschild
radii ($R_S$), the inclination of the accretion disk ($i$),
and the power-law of the radial emissivity dependence
($q$). The available statistics
is not sufficient to allow all these parameters to be
simultaneously constrained by the fit. The fits were therefore
performed in two steps. In the first
step, all the parameters were left free to
vary, and their confidence intervals calculated to determine
which of them are unconstrained. The unconstrained parameters
were then frozen to physically plausible values and/or
intervals in the second step. The
final results are summarized in Table~\ref{tab2}.
\begin{table*}
\caption{Best-fit results when the XMM-Newton/pn
iron line profile is fit with a relativistic profile}
\begin{center}
\begin{tabular}{lccccccc} \hline \hline
Observation & $E_c$ & $r_i$ & $r_o$ & $q$ & $i$ &
$EW$ & $\chi^2_{\nu}$ \\
& (keV) & ($R_S$) & ($R_S$) & & ($^{\circ}$) & (keV) & \\ \hline
Schwarzschild profile & $6.40^{+0.17}$ & $17 \pm^{<r_o}_8$ & $35 ^{+18}_{>
r_i}$ & $-2$$^a$ & $25 \pm^3_7$ & $340 \pm^{110}_{120}$ & 1.08 \\
Kerr profile & $6.40 ^{+0.05}$ & $8.2 \pm^{5.0}_{7.0}$ & $1700
\pm^{1700}_{800}$ & $-3$$^a$ & $< 39$ & $280 \pm^{110}_{760}$ & 1.09 \\
\hline \hline
\end{tabular}
\end{center}

\noindent
$^a$unconstrained and therefore fixed in the fit to the reported value

\label{tab2}
\end{table*}

\begin{figure}
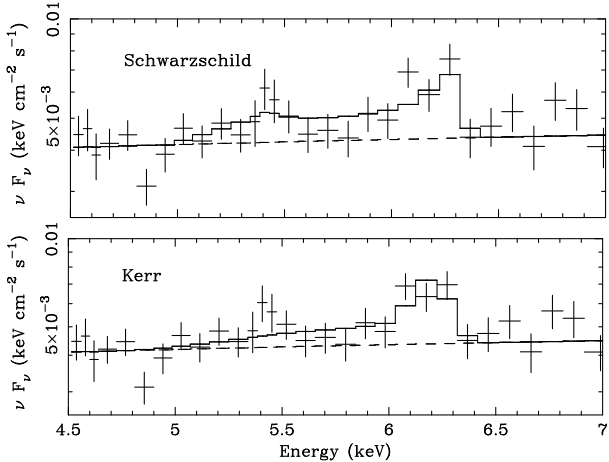

   \centering
   \includegraphics[angle=-90,width=8cm]{H3946_F7a.ps}
   \includegraphics[angle=-90,width=8cm]{H3946_F7b.ps}
      \caption{Spectrum ({\it crosses}) and best-fit
		model in the 4.5--7~keV energy
		band, when the iron K$_{\alpha}$
		emission line is modeled with a purely
		relativistic emission, around a
		Schwarzschild ({\it upper panel})
		and Kerr ({\it lower panel}) black
		hole
              }
         \label{fig9}
   \end{figure}

The comparison between the best-fit
relativistic profiles and the data is shown in Fig.~\ref{fig9}.
Deviations between individual data points and the
models are around $\pm 1$$\sigma$.
Nonetheless, a description of the
5--6.5~keV (observer's frame)
excess in terms of a double Gaussian
profile is better than the single relativistic
profile at the 94.6\% (Schwarzschild) and
96.5\% (Kerr) confidence level, respectively.
The relativistic profiles exhibit complementary
virtues and problems.
The Schwarzschild profile follows better
the overall excess shape, but remains slightly
below the observed blue peak; the Kerr profile
does not
``see'' the read peak, of course. Interestingly enough,
both models rules out
that the bulk of the photons are originally emitted
by He- or H-like iron, as $E_c \approxlt 6.6$~keV.
The addition of a farther narrow line component
underneath or besides the relativistically broadened
profile yields a negligible improvement
in the quality of the fit ($\Delta \chi^2 \approxlt 2$),
even if the
centroid energy of the narrow component is fixed to the value
corresponding to a fluorescent
neutral iron K$_{\alpha}$ line.

\begin{figure}
   \centering
   \includegraphics[angle=-90,width=8cm]{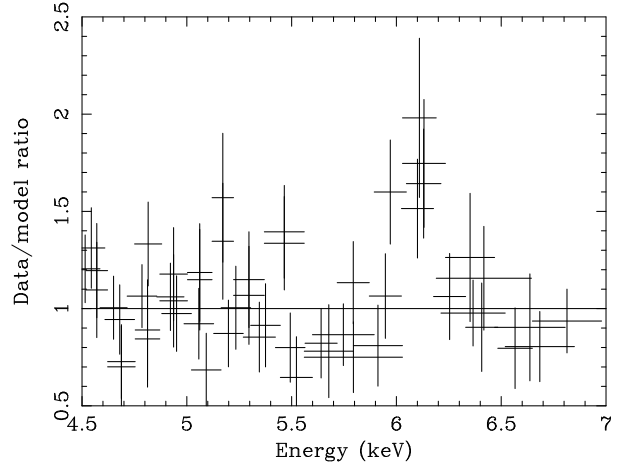}
      \caption{Residuals (in data/model ratio) when
	      a purely relativistic profile, with all
	      its parameter (except the normalization)
	     constrained within the best-fit XMM-Newton
             confidence intervals (cf. Table~\ref{tab2}),
	     is fit to the ASCA iron emission line
              }
         \label{fig11}
   \end{figure}

Limited constraints can be derived on the
iron line shape from either the ASCA or the
BeppoSAX observation. The ASCA line is clearly narrow
and no hint of a broad component exists.
Nonetheless, the upper limits
on the EW of an underlying relativistic
profile, with the same shape as observed by XMM-Newton,
are not particularly demanding: 720~eV and
300~eV for a Schwarzschild and a Kerr profile,
respectively (cf. Fig.~\ref{fig11}).
In the BeppoSAX spectrum,
due to the relative weakness of the
iron line, and to the comparatively
poorest energy resolution of the MECS detector,
the question whether the line profile is broad or narrow
remains basically unanswered.

\section{Discussion}

The history of the iron
K$_{\alpha}$ line profile in ESO198-G24 can
be summarized as follows (in order of increasing
2--10~keV flux):

\begin{itemize}

\item in the ASCA observation the line profile is
narrow, and no indication of a broad component
is present

\item the XMM-Newton pn spectrum
exhibits a broad line, centered at $E_c \simeq
6.4$~keV, alongside with a possible
(confidence level: 96.3\%) additional
feature at $E_c \simeq 5.7$~keV. The latter
feature does not correspond to any atomic transitions,
and would therefore
most likely belong to  a relativistic
profile together with the main ``6.4~keV" component

\item in the BeppoSAX observation the line has
an EW by a factor 2 (3) fainter than in XMM-Newton
(ASCA). It is impossible to
tell whether its profile is narrow or broad

\end{itemize}

If the emission feature at $E_c \simeq 5.7$~keV
detected in the pn spectrum
indeed belongs to a relativistic profile, the iron
line complex observed in ESO198-G24
may be one example of ``double-horned`` profile.
Unfortunately, the statistics in the rather short
XMM-Newton observation is not good enough to allow an
unambiguous description in terms of a
specific relativistically
broadened profile. Models employing the Schwarzschild
or Kerr kernel can equally well describe it
form the statistical point of view (although the
latter does not have any red peaks!).
Interestingly enough, the best description of the
excess in the 5--6.5~keV energy range
(observer's frame)
is in terms of a combination of two
Gaussians. The main reason for this is that
the 5.7~keV and 6.4~keV features
have comparable intensities (within a factor of 2),
while keeping a separation $\Delta E/E \simeq 0.15$.
They could hardly belong to a common
standard single relativistic profile, because such
a large relativistic broadening implies generally a
much stronger blue peak (\cite{fabian89}; \cite{matt91}).
Changes in the basic relativistic
profile model, such as absorption by inner shell
transitions of ionized iron (as proposed by
\cite{sako02}), or emission by a single
"flare", which illuminates only a small
fraction of the disk surface
(as proposed by \cite{nayakshin01} and
\cite{yaqoob01}), do not yield any
statistically significant improvements in
the quality of the fit.

The EW of the line observed by BeppoSAX
a few weeks later is by a factor of 2 weaker
than measured by XMM-Newton, whereas the
continuum flux was about 60\% more intense.
If the BeppoSAX profile is dominated
by a broad component, this evidence
suggests that the
response time of the relativistic line
to variations of the underlying continuum is probably
larger than at least a few hours.
If the solid angle subtended by the
reprocessing matter and the spatial distribution
of the intrinsic continuum did not change between
the two observations (which is consistent with
the closeness between the {\it intrinsic}
spectral index measured by BeppoSAX and the
{\it observed} spectral index measured
by XMM-Newton),
the larger EW of the relativistic
profile in the pn spectrum can be explained by
a delayed response to a brighter continuum flux
state before the start of the pn exposure. The
delay is at least of the order of the pn exposure
elapsed time, i.e. $\approxgt 2$~hours,
corresponding to a spatial scale
of $\approxgt 7 R_S M_8^{-1}$, if $M_8$ is the black hole
mass in units of $10^8$ solar masses.
Fabian et al. (2002) point out similar
problems to explain the short time scales
variability pattern
of the relativistic line in MCG-6-30-15.
If the BeppoSAX profile is substantially
"contaminated" by a narrow component, the discrepancy
between the broad component EWs is obviously even larger.

By contrast, the ASCA line profile is dominated
by a remarkably narrow "core".
Unresolved iron lines have
been now quite commonly discovered both in {\it Chandra}
grating (\cite{yaqoob01}, \cite{kaspi02})
and in XMM-Newton observations
(\cite{gondoin01a}; \cite{gondoin01b};
\cite{reeves01}; \cite{petrucci02}; \cite{obrien01}).
The two most plausible possibilities are
Compton-reflection by matter far off the central engine
(e.g., the ``torus''), or
gas clouds in the Broad Line Regions.
The EW of the narrow line in ASCA ($\simeq 300$~eV) is
much larger than typically observed 
(50--100~eV). If it originates in the torus,
one may expect $EW \simeq 100$~eV if the torus subtends
a solid angle
$\sim$$\pi$ (\cite{krolik94};
\cite{ghisellini94}). Such a large EW may be therefore
indicative 
either of a strong (a factor $\approxgt $10) iron
overabundance or, more likely, of the ``echo'' of 
a brighter illuminating flux state.
Our sparse knowledge of the historical X-ray light
curve of ESO198-G24 indeed suggests a dynamical
range of at least a factor 6 in the nuclear
power.

\section*{Appendix: On the "X-ray Baldwin effect"}

ESO198-G24 is one of
the brightest Seyfert~1s where
a fluorescent iron K$_{\alpha}$
line and a Compton reflection
continuum have been simultaneously measured.
The  amount of reflection (assuming an inclination angle
as derived by the fit of the
iron
line in the pn spectrum with a Schwarzschild profile:
$i = 25^{\circ}$) is 
$R = 0.4\pm^{0.4}_{0.3}$, about one-half the value expected if
the reprocessing occurs in a plane-parallel, semi-infinite
slab, and the primary emission is isotropic.

\begin{figure}
   \centering
   \includegraphics[width=8cm]{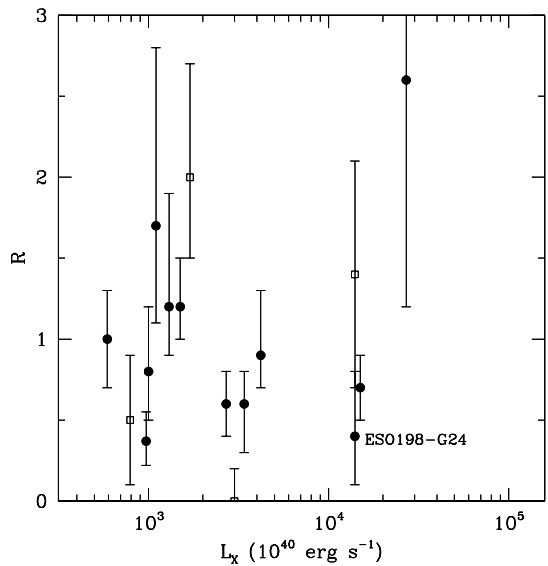}
      \caption{2--10~keV luminosity versus the Compton-reflection
		relative intensity $R$ for a PDS count rate
		limited sample of publicly available type
		1 ({\it filled circles}) and type~2
		({\it empty squares})
		Seyferts observed by BeppoSAX.
              }
         \label{fig13}
   \end{figure}

Nandra et al. (1997b) and Reeves \&
Turner (2000) discuss a possible ``X-ray Baldwin effect'',
whereby the intensity of the reprocessing features
decreases with increasing AGN X-ray luminosity due to
a higher degree of ionization of the accretion disk,
which smears the spectral contrast between the reflected
and direct continua and shifts the iron line centroid
either to intermediate ionization stages,
where
resonant scattering may cause
a reduction in the line flux (\cite{matt93}, 1996),
or to fully ionized stages. In ESO198-G24 there
is no compelling evidence for high iron ionization stages to be
responsible for the bulk of the line profile in any
of the scenarios discussed in this paper.
In order to test the dependence of the
Compton reflection on the luminosity,
we retrieved from the BeppoSAX public archive data
for a sample of Seyfert galaxies (of both type 1 and 2),
having a PDS count rate larger then $0.35$~s$^{-1}$
\footnote{The sample includes: ESO191-G55, Fairall~9, IC~4329A,
MCG~5-23-16, MCG~6-30-15,
MCG-8-11-11, Mkn~509, NGC~2110, NGC~3516, NGC~3783,
NGC~4151, NGC~4593, NGC~5506, NGC~5548, NGC~526A,
NGC~7469, . Measurements for MCG~6-30-15
are taken from \cite{guainazzi99b}, for NGC~4151
from \cite{schurch02}}. We
analyzed these observation, applying the standard
Seyfert spectral template, as described in Perola et al. (2002),
eventually modified for intervening photoelectric absorption
in type 2 objects. The reprocessor is assumed always
neutral, and the inclination angle
as in Sect.~3. The $R$ versus 2--10~keV luminosity
($L_X$) plot is shown in Fig.~\ref{fig13}.
There is no evidence for any correlations between these
quantities, across the 2.5 dex span in luminosity.

\begin{acknowledgements}

Comments by an anonymous referee strongly
contributed to improve the quality of the
presentation and sharpen the focus of this
paper.
This paper
is based on observations obtained with XMM-Newton, an ESA science
mission with instruments and contributions directly
funded by ESA Member States and the USA (NASA). The
XMM-Newton Science Archive (XSA) Development Team is
gratefully acknowledged for its highly professional
work.
This research has made use of data obtained through the
High Energy Astrophysics Science Archive Research Center
Online Service, provided by the
NASA/Goddard Space Flight Center
and of the NASA/IPAC Extragalactic Database (NED) which
is operated by the Jet Propulsion Laboratory,
California Institute of Technology,
under contract with the National Aeronautics and Space
Administration. 

\end{acknowledgements}

\end{document}